\documentclass[5p]{elsarticle}

\usepackage{amssymb}
\usepackage{mathpazo}
\usepackage{booktabs}
\usepackage{color}
\usepackage{units}
\usepackage{amstext}
\usepackage{graphicx}
\usepackage{amsmath}
\usepackage{esint}
\usepackage[colorlinks,citecolor=blue,filecolor=black,linkcolor=red,urlcolor=blue]{hyperref}
\usepackage{breakurl}

\makeatletter
\usepackage{color}
\biboptions{sort&compress}
\usepackage{breakurl}
\begin{document}

\begin{frontmatter}




\title{Solitons supported by a self-defocusing trap in a fractional-diffraction waveguide}


\author[inst1]{Mateus C. P. dos Santos\corref{cor1}}
\ead{mateuscalixtopereira@gmail.com}
\affiliation[inst1]{organization={Instituto de Física, Universidade Federal de Goiás},%Department and Organization
            addressline={74.690-970}, 
            city={Goiânia},
            %postcode={00000}, 
            state={Goiás},
            country={Brazil}}

\affiliation[inst2]{organization={Department of Physical Electronics, School of Electrical Engineering, Faculty of Engineering, and Center for Light-Matter Interaction, Tel Aviv University},%Department and Organization
            addressline={P.O.B. 39040}, 
            city={Tel Aviv},
            %postcode={00000}, 
            %state={Goiás},
            country={Israel}}
\author[inst2,inst3]{Boris A. Malomed}
\ead{malomed@tauex.tau.ac.il}

\author[inst1]{Wesley B. Cardoso}
\ead{wesleybcardoso@ufg.br}

\affiliation[inst3]{organization={Instituto de Alta Investigación, Universidad de Tarapacá},%Department and Organization
            addressline={Casilla 7D}, 
            city={Arica},
            %postcode={22222}, 
            %state={State Two},
            country={Chile}}

\cortext[cor1]{Corresponding author}

\begin{abstract}
We introduce a model which gives rise to self-trapping of fundamental and
higher-order localized states in a one-dimensional nonlinear Schr\"{o}dinger
equation with fractional diffraction and the strength of the self-defocusing
nonlinearity growing steeply enough from the center to periphery. The model
can be implemented in a planar optical waveguide. Stability regions are
identified for the fundamental and dipole (single-node) states in the plane
of the L\'{e}vy index and the total power (norm), while states of higher
orders are unstable. Evolution of unstable states is investigated too,
leading to spontaneous conversion towards stable modes with fewer node.
\end{abstract}

%Graphical abstract



\end{frontmatter}


\section{Introduction}

The fractional Schr\"{o}dinger equation (FSE) was first introduced by N.
Laskin for wave function $\Psi $ of a quantum particle moving by L\'{e}vy
flights. \cite{Lask1,Laskin_PRA02,Lask2}. In the scaled form, with time $t$
and coordinate $x$, the one-dimensional FSE derived in those works is (see
also Ref. \cite{Hu_ANO00})
\begin{equation}
i\frac{\partial \Psi }{\partial t}=\frac{1}{2}\left( -\frac{\partial ^{2}}{%
\partial x^{2}}\right) ^{\alpha /2}\Psi +V(x)\Psi ,  \label{FSE}
\end{equation}%
where $V(x)$ is the usual potential which appears in the usual Schr\"{o}%
dinger equation. The kinetic-energy operator in Eq. (\ref{FSE}), with the L%
\'{e}vy index (LI)\ $\alpha $, which may take values
\begin{equation}
0<\alpha \leq 2,  \label{interval}
\end{equation}%
\cite{Benoit}, is represented by the fractional \textit{Riesz derivative}
\cite{Riesz1,Riesz2}, which is defined as the juxtaposition of the direct
and inverse Fourier transforms \cite{Jeng_JMP10,Luchko_JMP13,Duo_CMA16},
\begin{equation}
\left( -\frac{\partial ^{2}}{\partial x^{2}}\right) ^{\alpha /2}\Psi =\frac{1%
}{2\pi }\int_{-\infty }^{+\infty }dp|p|^{\alpha }\int_{-\infty }^{+\infty
}d\xi e^{ip(x-\xi )}\Psi (\xi ),  \label{Riesz derivative}
\end{equation}%
where $p$ is the Fourier-space wavenumber conjugate to coordinate $x$. The
limit value $\alpha =2$ corresponds to the usual (non-fractional)
kinetic-energy operator, $-\partial ^{2}/\partial x^{2}$.

Proceeding from the single-particle FSE (\ref{FSE}), one may attempt to
introduce its generalization in the form of the Gross-Pitaevskii equation
\cite{Pit-Str} for the Bose-Einstein condensate in an ultracold gas of L\'{e}%
vy-flying particles:%
\begin{equation}
i\frac{\partial \Psi }{\partial t}=\frac{1}{2}\left( -\frac{\partial ^{2}}{%
\partial x^{2}}\right) ^{\alpha /2}\Psi +V(x)\Psi +\sigma |\Psi |^{2}\Psi ,
\label{FGPE}
\end{equation}%
where $\sigma =+1$ or $-1$ corresponds to repulsive or attractive contact
interactions, respectively, between the particles \cite{review,HS}. However,
a consisted derivation of the fractional Gross-Pitaevskii equation in the
framework remains a challenging problem.

Realizations of the fractional quantum mechanics were also proposed in
solid-state L\'{e}vy crystals \cite{Levycrystal} and exciton-polariton
condensates in semiconductor microcavities \cite{Pinsker}. However, no
experimental demonstration of fractional quantum systems has been
demonstrated, as yet.

A more feasible approach to the physical realization of FSEs in the form of
classical equations is suggested by the commonly known similarity of the Schr%
\"{o}dinger equation in quantum mechanics and the propagation equation for
the optical field under the condition of paraxial diffraction, with time $t$
replaced in Eq. (\ref{FSE}) by the propagation distance, $z$, which plays
the role of the evolution variable in optical waveguides \cite{KA}. A scheme
for the realization of this possibility was proposed by Longhi \cite%
{Longhi_OL15}, in terms of the transverse light dynamics in optical cavities
with the $4f$ (four-focal-lengths) structure. The proposed setup
incorporates two lenses and a phase mask, which is placed in the middle
(Fourier) plane. The first lens performs decomposition of the light beam
into Fourier components, with respect to the transverse coordinates, while
the second lens recombines the component back into a single beam. The action
of the fractional diffraction is emulated, in the Fourier plane, by the
corresponding differential phase shifts imposed by the phase mask. The
scheme may also include a curved mirror placed at an edge of the cavity,
which introduces a phase shift representing potential $V\left(x\right)$ in
Eq. (\ref{FSE}).

Experimental implementation of such an optical setup has not been reported
as yet. However, effective fractional group-velocity dispersion has been
recently realized in a fiber-laser scheme \cite{Shilong}.

Once the FSE may be implemented as the propagation equation in optics, a
natural possibility is to include the self-focusing or defocusing
nonlinearity of the optical material. The respective fractional nonlinear
Schr\"{o}dinger equation (FNLSE) for amplitude $U(x,z)$ of the
electromagnetic wave is \citep{review,Zeng2019,Chen_CHAOS20}
\begin{equation}
i\frac{\partial}{\partial z}U=\frac{1}{2}\left( -\frac{\partial ^{2}}{%
\partial x^{2}}\right) ^{\alpha /2}U+\eta |U|^{2}U,  \label{FNLSE}
\end{equation}%
where $\eta $ is the nonlinearity coefficient, and the external potential is
neglected. The regular (non-fractional) diffraction corresponds to $\alpha
=2 $ \citep{Laskin_PRA02,Hu_ANO00}. In the case of self-focusing, which
corresponds to $\eta <0$ in Eq. (\ref{FNLSE}), the FNLSE (\ref{FNLSE}) gives
rise to a family of \textit{fractional solitons }\cite%
{Secchi,Chen,Dong1,Dong2,Dong3}, see also review \cite{review}. They exist
in the interval of $1<\alpha \leq 2$, as the same FNLSE gives rise to the
\textit{collapse} at $\alpha <1$ \cite{review}. On the other hand, the
collapse does not take place in the case of self-defocusing, which
corresponds to $\eta >0$ in Eq. (\ref{FNLSE}) and is the subject of the
present work. Therefore, we here consider the full interval (\ref{interval}%
). Recently, nonlinear modes in the form of \textit{domain walls} were
studied in a system of two coupled FNLSEs with the self-repulsive
nonlinearity, in the full interval of $0<\alpha \leq 2$ \cite{Strunin}.

A possibility to produce soliton-like modes in nonlinear Schr\"{o}dinger
equations with the defocusing nonlinearity was proposed in Refs. \cite%
{Borovkova_PRE11,Borovkova_OE12,ZENG_MALOMED_PRE12,Tian,Wu,Cardoso_PRE13,Raymond}%
. It makes use of spatial modulation of the nonlinearity strength, growing
steeply enough from the center to periphery, which gives rise to
self-trapping of various one- and two-dimensional localized modes. The
objective of the work is to investigate this possibility in the
one-dimensional setting with the fractional diffraction. The model is
introduced in Section 2. Results of systematic numerical investigation of
fundamental and higher-order self-trapped states are summarized in Section
3, the most important results concerning stability of these states. The
paper is concluded by Section 4.

\section{The model}

We consider an optical beam propagating in a planar waveguide with
self-defocusing nonlinearity and fractional diffraction. The evolution of
amplitude $U(z,t)$ of the optical field is governed by the FNLSE. In the
rescaled form, it takes the form,
\begin{equation}
i\frac{\partial }{\partial z} U=\frac{1}{2}\left( -\frac{\partial ^{2}}{%
\partial x^{2}}\right) ^{\alpha /2}U+\eta (x)|U|^{2}U,  \label{EQ1}
\end{equation}%
where $z$ is the normalized propagation distance and $x$ the transverse
coordinate. The fractional diffraction with LI $\alpha $
\cite{Benoit} is represented in Eq. (\ref{EQ1}) by the Riesz derivative \cite%
{Riesz1,Riesz2}. It is defined, in terms of the Fourier-transform operator $%
\mathcal{F}$, as \citep{Jeng_JMP10,Luchko_JMP13,Duo_CMA16}
\begin{equation}
\left( -\frac{\partial ^{2}}{\partial x^{2}}\right) ^{\alpha /2}U(x,z)=%
\mathcal{F}^{-1}\left\{ |\omega |^{\alpha }\mathcal{F}[U(x,z)]\right\} ,
\label{FRAC}
\end{equation}%
where $\omega $ is the wavenumber in the Fourier space conjugate to
coordinate $x$.

Here, we assume that the self-trapping of localized states is provided by
local modulation of the defocusing cubic nonlinearity, defined in Eq. (\ref%
{EQ1}) by coefficient $\eta (x)>0$ with a minimum at $x=0$, which grows
steeply enough at $|x|\rightarrow \infty $. The creation of various stable
soliton-like modes by means of the nonlinear potential well induced by the
appropriate profile of $\eta (x)$ in the case of the regular diffraction ($%
\alpha =2$) was elaborated in Refs. %
\citep{Borovkova_PRE11,Borovkova_OE12,ZENG_MALOMED_PRE12,Cardoso_PRE13}.
Such modes with propagation constant $b<0$ are looked for as%
\begin{equation}
U(x,z)=e^{ibz}u(x),  \label{Uu}
\end{equation}%
with real function $u$ satisfying the equation%
\begin{equation}
bu+\frac{1}{2}\left( -\frac{\partial ^{2}}{\partial x^{2}}\right) ^{\alpha/2}u+\eta (x)u^{3}=0.  \label{EQ2}
\end{equation}%

\begin{figure}[t!]
\centering \includegraphics[width=0.95\columnwidth]{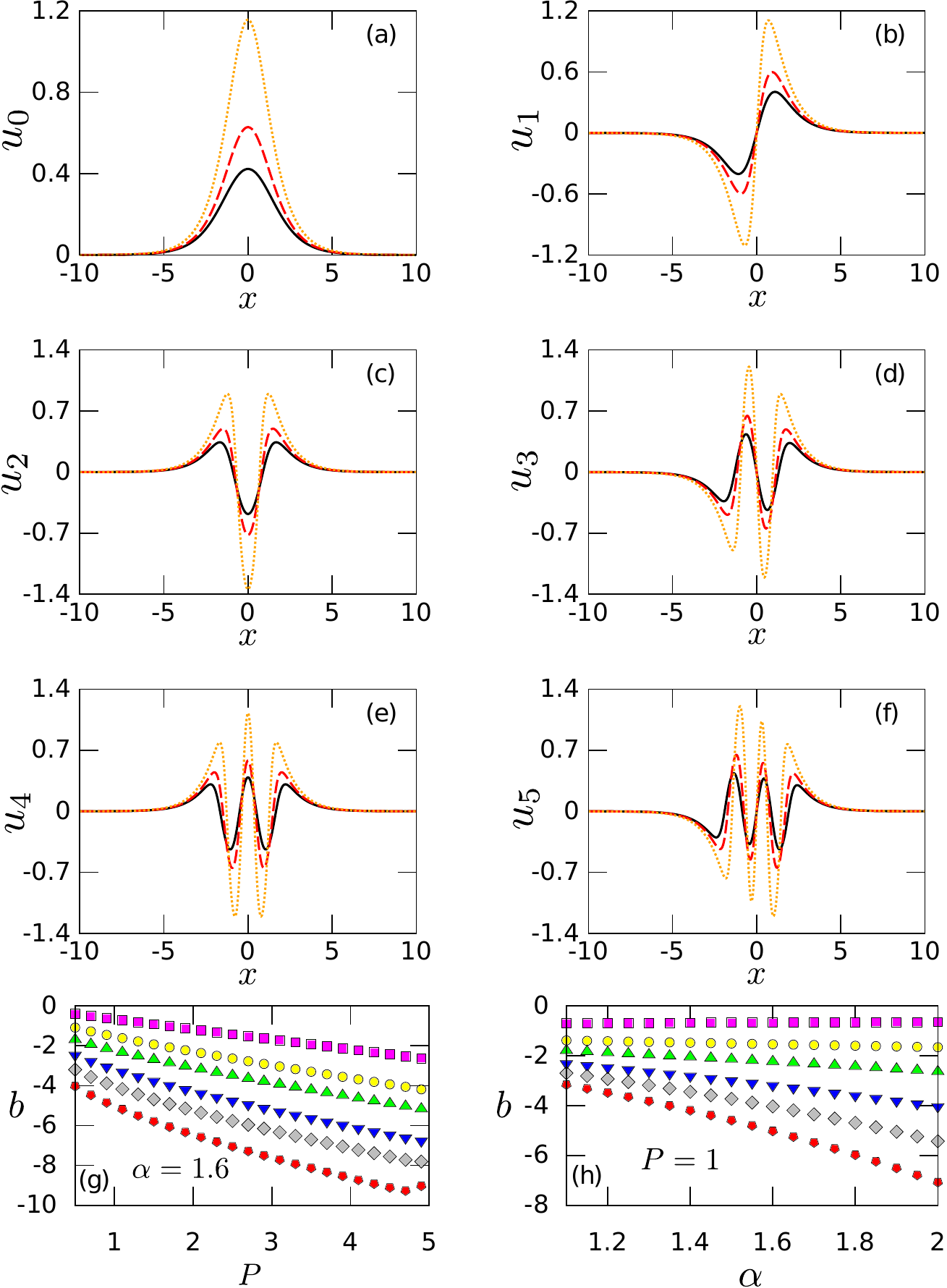}
\caption{Profiles of bound states: (a) $u_{0}$, (b) $u_{1}$, (c) $u_{2}$,
(d) $u_{3}$, (e) $u_{4}$, and (f) $u_{5}$, as produced by the numerical
solution of Eq. (\protect\ref{EQ2}), with LI $\protect\alpha =1.6$, based on
the relaxation method. The profiles with powers $P=0.5$, $1$ and $3$ are
displayed by solid (black), dashed (red) and dotted (orange) lines,
respectively. Panels (g) and (h) display, severally, the propagation
constant $b$ vs. $P$ for fixed $\protect\alpha =1.6$, and $b$ vs. $\protect%
\alpha $ for fixed $P=1$, by means of squares (magenta), circles (yellow),
triangles (green), inverted triangles (blue), diamonds (gray), and hexagons
(red), which correspond to the states $u_{0}$, $u_{1}$, $u_{2}$, $u_{3}$, $%
u_{4}$ and $u_{5}$, respectively.}
\label{F1}
\end{figure}
The asymptotic form of the solution to Eq. (\ref{EQ2}) at $|x|\rightarrow
\infty $ is correctly produced by the Thomas-Fermi approximation, which
neglects the diffraction term in the equation \cite{Peli},%
\begin{equation}
u_{\mathrm{TF}}^{2}(x)=-b/\eta (x).  \label{TF}
\end{equation}%
As it follows from Eq. (\ref{TF}), the condition necessary for the
convergence of the soliton's total power (norm), which is a dynamical
invariant of Eq. (\ref{EQ1}),%
\begin{equation}
P=\int_{-\infty }^{+\infty }u^{2}(x)dx,  \label{NORM}
\end{equation}%
is that $\eta (x)$ grows at $|x|\rightarrow \infty $ faster than $|x|$. In
most works %
\citep{Borovkova_PRE11,Borovkova_OE12,ZENG_MALOMED_PRE12,Cardoso_PRE13}, the
modulation is adopted in a steeper form -- in particular, as
\begin{equation}
\eta (x)=\cosh ^{2}(x),  \label{cosh^2}
\end{equation}%
which we adopt here. In optics, such modulation formats may be realized by
means of the spatially inhomogeneous distribution of dopants resonantly
interacting with the light beam, or by a spatially inhomogeneous profile of
the resonance detuning in the dopants \citep{Kip,Kartashov_RMP11}.

The stability of static solutions (\ref{Uu}) is addressed by means of
linearized equations for small perturbations, taking a solution as
\begin{equation}
U_{\text{p}}(x,z)=e^{ibz}\left[u(x)+v(x)e^{\lambda z}+w^{\ast}(x)e^{\lambda^{\ast}z}\right],  \label{pert}
\end{equation}%
where $v(x)$ and $w(x)$ are components of the eigenmode of the perturbation
and $\lambda $ is an eigenvalue (that may be complex), the stability
condition being Re$(\lambda )=0$ for all eigenvalues. To substitution of
expression (\ref{pert}) $U_{\text{p}}$ into Eq. (\ref{EQ1}) and
linearization with respect to $v(x)$ and $w(x)$ leads to the eigenvalue
problem,
\begin{equation}
-i\left[
\begin{array}{cc}
C_{1} & C_{2} \\
-C_{2} & -C_{1}%
\end{array}%
\right] \left[
\begin{array}{c}
v \\
w%
\end{array}%
\right] =\lambda \left[
\begin{array}{c}
v \\
w%
\end{array}%
\right] ,  \label{STAB_LIN}
\end{equation}%
where we define
\begin{equation}
C_{1}\equiv \frac{1}{2}\left( -\frac{\partial ^{2}}{\partial x^{2}}\right)
^{\alpha /2}+b+2\eta u^{2},  \label{c1}
\end{equation}%
\begin{equation}
C_{2}\equiv \eta u^{2}.  \label{c2}
\end{equation}%
The problem based on Eq. (\ref{STAB_LIN}) can be solved by means of the
Fourier collocation method \citep{Yang_10}.

\section{Numerical results \label{Sec3}}

The ground-state and higher-order (excited) stationary solutions of Eq. (\ref%
{EQ1}), $u_{n}$, with $n=0,1,2,...$, where $n$ is the number of nodes in $%
u_{n}(z)$, were found by means of the well-known imaginary-time evolution
method \cite{imaginary}. This procedure was implemented using the split-step algorithm based on the Fourier spectral method with numerical steps $\Delta_x=0.04$ and $\Delta_z=0.001$. Hermite-Gaussian functions were employed as the natural input profiles. In the numerical calculations, the
higher-order modes, with $n\geq 1$, are obtained using the Gram-Schmidt
orthogonalization process, performed at each integration step %
\citep{Cardoso_PRE13}. Alternatively, all these modes can be obtained by
means of the relaxation method, using the same input profiles.

To study the stability and dynamics of the stationary states, we calculated
the spectrum of perturbation eigenvalues and verified the corresponding
predictions for the stability, running real-time simulations of the
perturbed evolution.

First, examples of the family of solutions with $n=0,1,2,3,4$ and $5$, for
LI $\alpha =1.6$ and powers $P=0.5,1$ and $3$ [see Eq. (\ref{NORM})], are
displayed in Figs. \ref{F1}(a)-\ref{F1}(f), where we observe that the
increase in the power naturally produces solutions with larger amplitudes,
maintaining the number of nodes. In Figs. \ref{F1}(g) and \ref{F1}(h) the
results are summarized by dependences $b(P)$ for fixed $\alpha =1.6$ and $%
b(\alpha )$ for fixed $P=1$. We observe, in particular, that all families of
stationary states satisfy the \textit{anti-Vakhitov-Kolokolov} criterion, $%
db/dP<0$. It is a necessary condition for the stability of localized states
supported by a self-defocusing nonlinear term \cite{anti} (the
Vakhitov-Kolokolov criterion per se, $db/dP>0$, is a well-known stability
condition for solitons in the case of self-focusing \cite{VK,Berge,Fibich}).
On the other hand, the lowest modes $u_{n}$, with $n=0$ and $1$, demonstrate
a very weak dependence of their propagation constants on LI, the dependence
getting conspicuous ($db/d\alpha <0$) for $n\geq 2$. This finding is
explained by the fact that the above-mentioned TF approximation, which
ignores the derivative term in Eq. (\ref{EQ1}), plays a dominant role for
the modes with the simplest structure, while more sophisticated ones, with a
larger number of nodes, are sensitive to the diffraction effect, i.e., to $%
\alpha $. The negative sign of $db/d\alpha $ is explained by the fact that,
around the peaks in the shape of $\left\vert u_{n}(x)\right\vert $, the sign
of the curvature is opposite to the sign of $u_{n}(x)$ of the trapped state,
therefore the diffraction term makes a negative contribution to $b$, with
the size growing with the increase of $P$.

\begin{figure}[tb]
\centering \includegraphics[width=0.95\columnwidth]{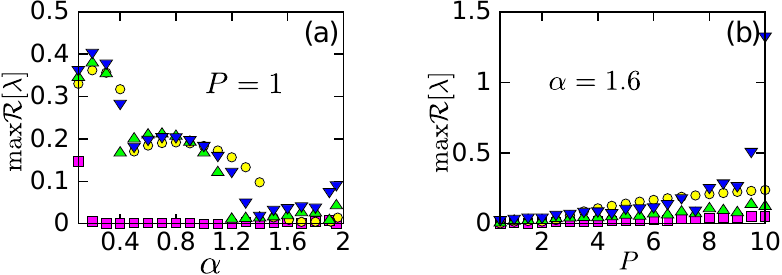}
\caption{The maximum value of the instability growth rate, Re$(\protect%
\lambda )$, for the bound states, as a function of $\protect\alpha $ (a) and
$P$ (b). Re$(\protect\lambda )$ is produced by the numerical solution of Eq.
(\protect\ref{STAB_LIN}). The results for the bound states $u_{0}$, $u_{1}$,
$u_{2}$, and $u_{3}$ are presented by squares (magenta), circles (yellow),
triangles (green), and inverted triangles (blue), respectively.}
\label{F2}
\end{figure}

Proceeding to the stability analysis of the bound states, in Figs. \ref{F2}%
(a) and \ref{F2}(b) we plot, severally, largest real parts of the stability
eigenvalues produced by the numerical solution of Eq. (\ref{STAB_LIN}), as
functions of $\alpha $ for $P=1$, and as functions of $P$ for fixed $\alpha
=1.6$, for the self-trapped modes $u_{n}$ with $n=0,1,2$ and $3$. We observe
that the nodeless ground-state modes, $u_{0}$, with $P=1$ are completely
stable for $\alpha >0.28$. This is not the same for the higher-order modes.
The one with $n=1$ features a restricted stability region, $\alpha >\alpha
_{\min }^{(n=1)}\approx 1.61$, while the modes $n>1$ are completely unstable
at all values of $\alpha <2$.

Another essential result is the analysis of the stability as a function of
the power, for a fixed value of LI. Indeed, in Fig. \ref{F2}(b), we observe that the increase in $P$ leads to destabilization of modes. For the case with $\alpha =1.6$, as shown in Fig. 2(b), the self-trapped nodeless mode is not stable for all values of $P$, presenting instability for $P > 2.8$. The same behavior is observed for the mode
$u_{1}$. Similar to the case shown in Fig. \ref{F2}(a), no stability region
is found for modes with $n>1$ (although it is seen in Fig. \ref{F2}(a) that
the instability of the states with $n=2$ is very weak). The latter
conclusion pertains as well to the case of the normal (non-fractional)
diffraction, $\alpha =2$. In this connection, it is relevant to mention
that, in the case of the still steeper \textit{anti-Gaussian} modulation of
the local self-defocusing strength, rather than the profile (\ref{cosh^2})
adopted here, the state with $n=2$ is stable too at $\alpha =2,$ while the
instability commences from $n=3$. The complex behavior of the stability region of the bound states, considering the variations of $\alpha$ and $P$, is studied in detail below, and the results are shown in Fig. \ref{F6}.

To verify the predictions for the stability and instability of the bound
states produced by the linear-stability analysis, we have performed direct
simulations using as input the numerically obtained stationary solutions
with amplitudes randomly perturbed at the $\pm 3$\% level. Absorbing
boundary conditions were applied at edges of the integration domain to
preclude artifacts produced by reflected radiation waves. The numerical
solution performed with the edge absorbers may break the conservation of the
power, defined by Eq. (\ref{NORM}), which would be one of indicators for the
instability of the underlying stationary solution.

\begin{figure}[t!]
\centering \includegraphics[width=0.95\columnwidth]{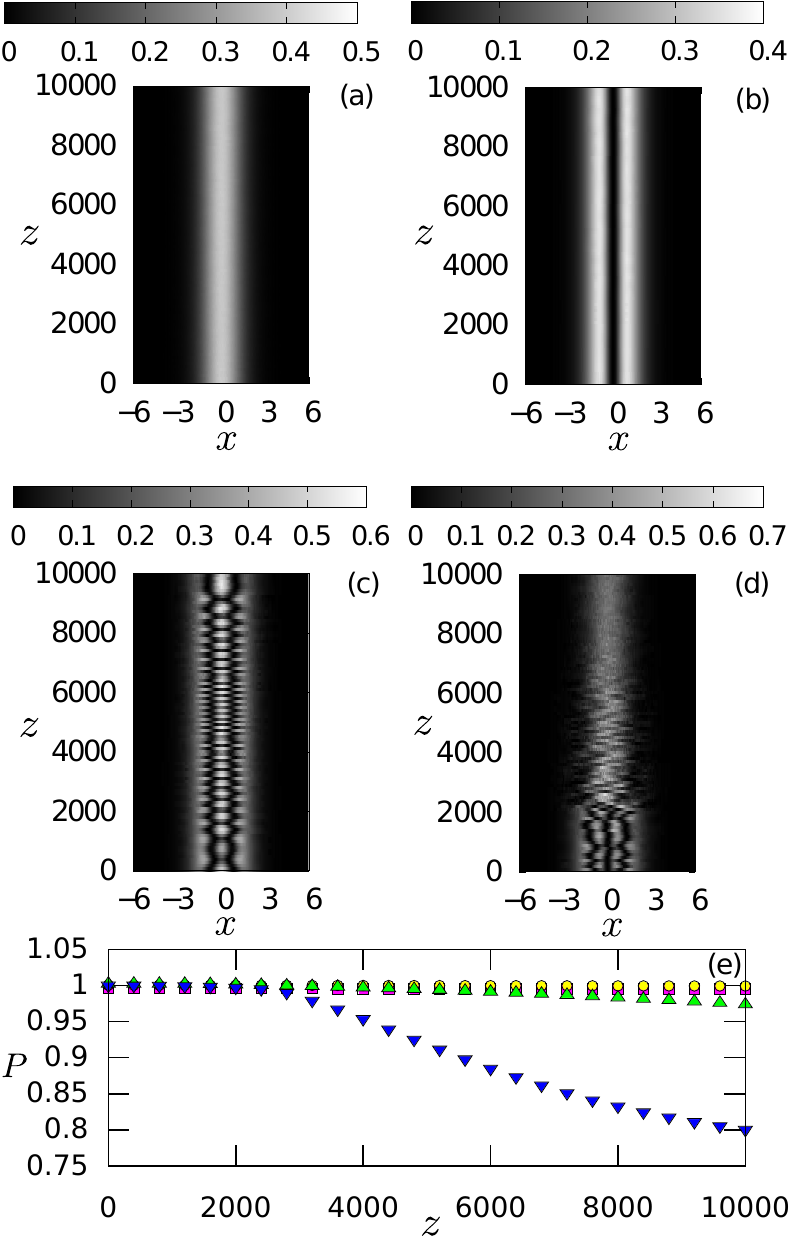}
\caption{The evolution of the perturbed modes with $\protect\alpha =1.7$ and
$P=1$. The results are presented for (a) $|u_{0}|^{2}$, (b) $|u_{1}|^{2}$,
(c) $|u_{2}|^{2}$, and (d) $|u_{3}|^{2}$. Panel (e) displays the total power
versus $z$ for the cases shown in panels (a-d), displayed by squares
(magenta), circles (yellow), triangles (green), and inverted triangles (blue) for $u_{0}$, $u_{1}$, $u_{2}$ and $u_{3}$, respectively.}
\label{F3}
\end{figure}

Figure \ref{F3} displays the perturbed evolution of the bound states $u_{0}$%
, $u_{1}$, $u_{2}$ and $u_{3}$ for $\alpha =1.7$ and $P=1$. It is seen that $%
u_{0}$ and $u_{1}$ demonstrate full dynamical stability. On the other hand,
modes $u_{n}$ with $n\geq 2$ exhibit instabilities. In particular, $u_{2}$
displays nonperiodic oscillations of its amplitude. The evolution of $u_{3}$
shows a more dramatic unstable behavior, in which the initial profile with
three nodes abruptly transforms into the zero-node one. In all the cases,
the results are in agreement with the stability/instability predictions
produced by the eigenvalues of small perturbations, cf. see Fig. \ref{F2}.

The evolution of the total power corresponding to the same cases is shown in
Fig. \ref{F3}(e). The power of the stable states ($u_{0,1}$) remains
constant, while the unstable configurations demonstrate losses, due to the
above-mentioned effect of the elimination of developing perturbations by the
boundary absorbers. In particular, mode $u_{3}$ loses $20$\% of the initial
power loss by $z=10^{4}$. Eventually, the remaining state converges to a
nodeless single-peak ground state, which keeps $\approx 70$\% of the initial
power. In general, higher-order unstable modes with larger values of $\alpha
$ (stronger diffraction) tend to show more abrupt power loss which commences
earlier.

\begin{figure}[t!]
\centering \includegraphics[width=0.95\columnwidth]{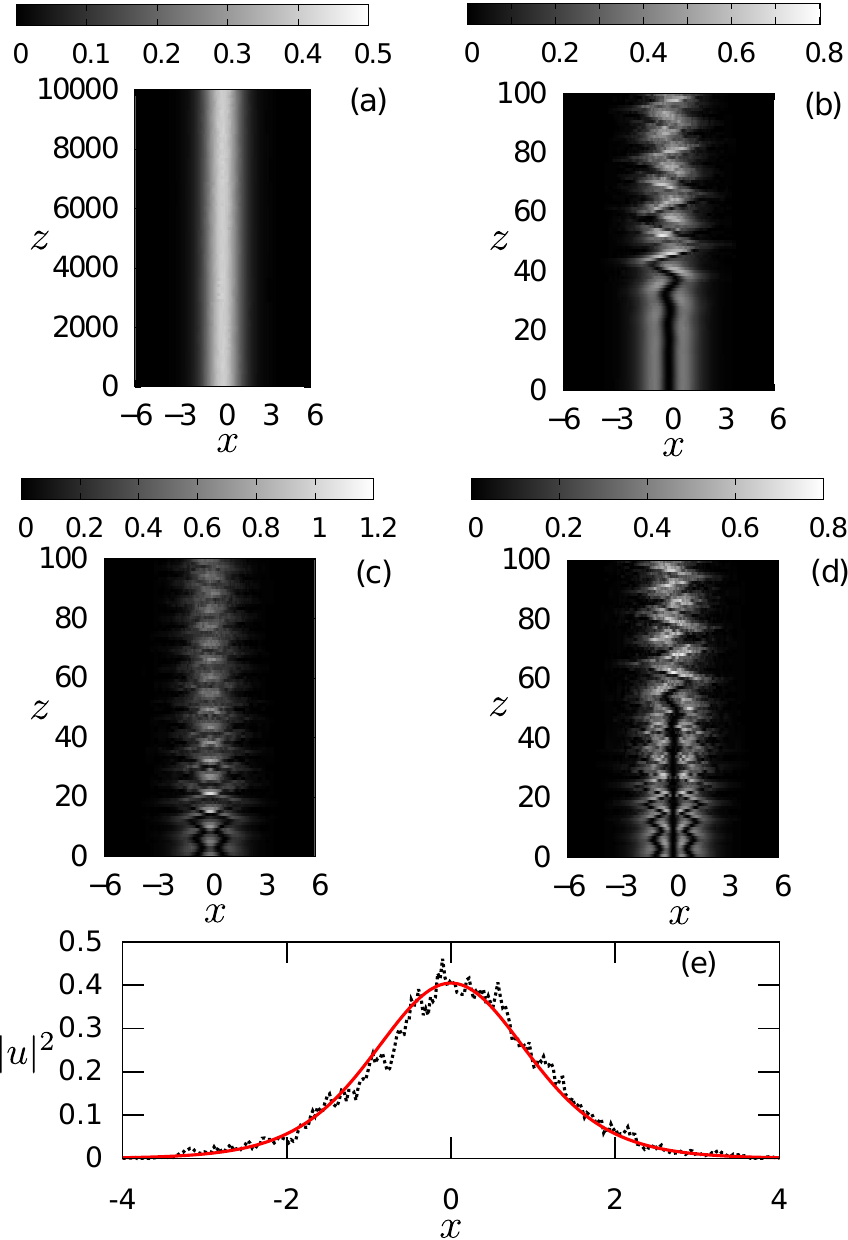}
\caption{Panels (a)-(d) display the same as in Figs. \protect\ref{F3}(a)-%
\protect\ref{F3}(d), but for $\protect\alpha =1.2$. (e) Comparison between
the intensity of the nodeless ground state $|u_{0}|^{2}$ (the solid line)
and the output produced by the perturbed evolution (spontaneous conversion)
of the first-order mode $|u_{1}|^{2}$ at $z=9600$ (the dotted line). }
\label{F4}
\end{figure}

The effect of LI on the stability of the bound states is presented in Fig. %
\ref{F4}, which displays the evolution of the states similar to those in
Fig. \ref{F3}, but with a smaller LI, $\alpha =1.2$. In this case, the
nodeless ground state remains completely stable. On the other hand, the $%
u_{1}$ mode is now unstable, in contrast with its counterpart shown above
for $\alpha =1.7$. The instability development abruptly transforms its shape
from the single-node one into a robust nodeless one. This result may be
identified as spontaneous transformation of the unstable $u_{1}$ state into
the stable ground state $u_{0}$, as shown in Fig. \ref{F4}(e). In this case (%
$\alpha =1.2$), all excited states $u_{n}$ with $n\geq 1$ are unstable,
indicating that the decrease of $\alpha $, i.e., weaker fractional
diffraction, leads to destabilization of the excited states. In some cases,
the instability triggers a more complex spontaneous evolution. For example,
unstable mode $u_{3}$ transforms first to $u_{2}$, before converging later
to the nodeless mode. On the other hand, the instability-driven evolution in
the case of lower values $\alpha $ does not exhibit conspicuous power loss,
i.e., the weaker fractional diffraction gives rise to very weak emission of
radiation.

\begin{figure}[t!]
\centering \includegraphics[width=0.95\columnwidth]{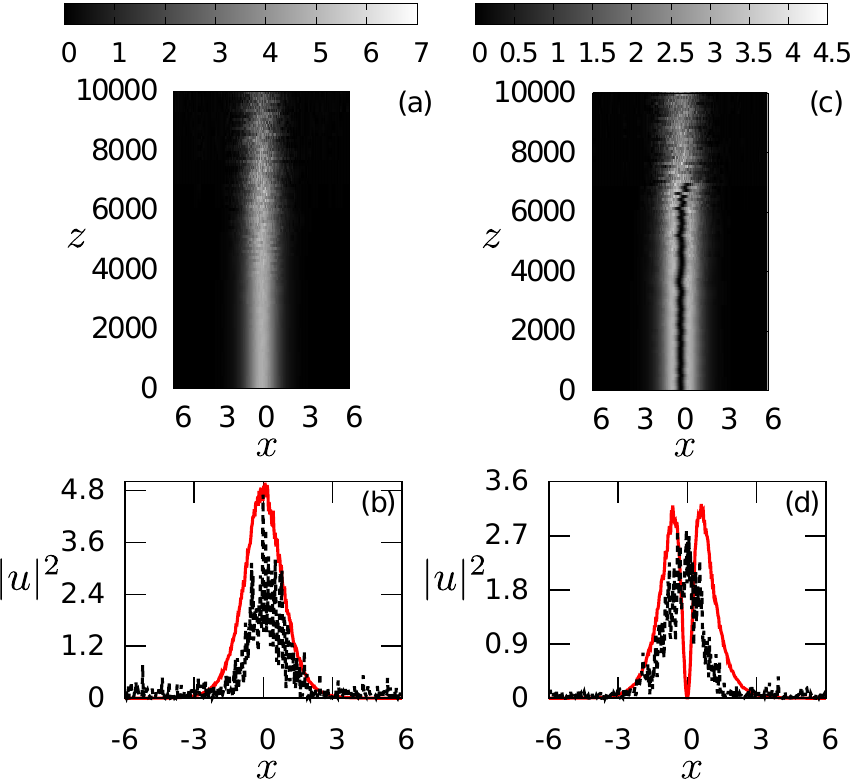}
\caption{Upper panels show the unstable evolution of the modes with $\protect%
\alpha =1.7$, while the lower panels display the comparison between the
density of the initial state at $z=0$ (solid lines) and in the output at $%
z=10000$ (dotted lines). The initial powers are $P=10$ for the nodeless mode
$u_{0}$ in (a,b), and $P=7$ for the first excited mode $u_{1}$ in (c,d).}
\label{F5}
\end{figure}

\begin{figure*}[t!]
\centering \includegraphics[width=0.8\textwidth]{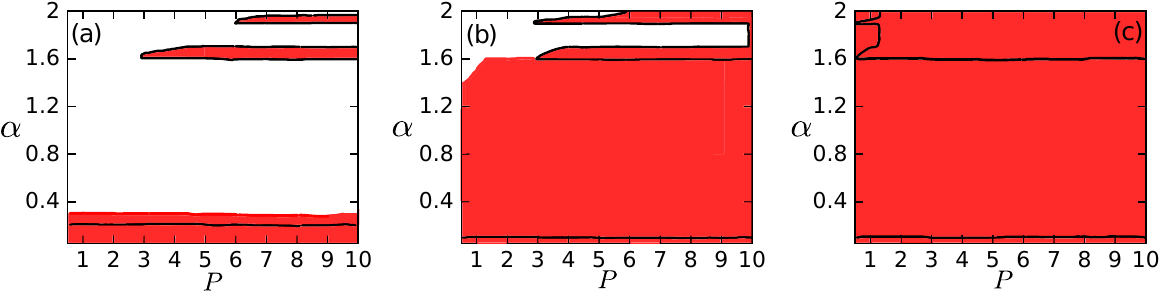}
\caption{Stability charts for modes $u_{0}$ (a), $u_{1}$ (b) and $u_{2}$
(c). The solutions are stable and unstable in white and red regions,
respectively. The areas delimited by the solid black line indicate
instability regions where the solutions spontaneously transform into others
with fewer nodes (eventually, into nodeless ones) and a lower norm.}
\label{F6}
\end{figure*}

We have also verified how the perturbed evolution of modes depends on total
power, $P$, confirming the predictions of the linear-stability analysis
indicate that the increase in $P$ should lead to destabilization. A example
of the evolution of modes $u_{0}$ and $u_{1}$ is displayed in Fig. \ref{F5},
for the same configurations as in Fig. \ref{F3}, but with high values of $P$%
. In particular, the unstable evolution of the nodeless mode is observed at $%
\alpha =1.7$ and $P=10$. In this case, the evolution of the unstable
nodeless mode leads to a power loss of $28.9$\% by $z=10^{4}$, induced by
the boundary absorbers. The evolution and comparison between the input and
output profiles are shown in Figs. \ref{F5}(a,b). Somewhat similar evolution
of the first-order mode $u_{1}$ is reported in Figs. \ref{F5}(c,d), for $%
\alpha =1.7$ and $P=7$. The evolution of this input produces a $24.3$\%
power loss and ends up by the spontaneous transformation into a nodeless
single-shape profile. A general trend is towards rearrangement of the
unstable nodeless state into a stable one with a smaller power.

Finally, Fig. \ref{F6} summarizes results of this work in the form of
stability charts for modes $u_{0}$, $u_{1}$ and $u_{2}$ in the parameter
plane ($P$, $\alpha $). We observe that the nodeless state is fully stable,
as may be expected from the ground state, for all values of the power in a
broad interval of the LI values, $0.3<\alpha <1.6$, as well as in a narrow
one,
\begin{equation}
1.7<\alpha <1.9.  \label{stable}
\end{equation}%
When the nodeless modes are unstable, they tend to evolve with with power losses, as
shown above. The dynamics of the first excited mode, $u_{1}$, are quite
different, but they share the stability interval (\ref{stable}) with the
nodeless mode. Finally, no stability region was found
for the second-order mode, $u_{2}$.

\section{Conclusion \label{Sec4}}

We have demonstrates that the FNLSE (fractional nonlinear Schr\"{o}dinger
equation) with the trapping nonlinear potential, induced by the spatially
modulated defocusing cubic nonlinearity, maintains stable self-trapped
modes. The model can be realized in a nonlinear planar waveguide with the
effective fractional diffraction. The linear-stability analysis and direct
simulations show that the nodeless modes, as well as the dipole single-node
ones (the first excited state) have well-defined stability regions in the
plane of the LI (L\'{e}vy index) and power. On the other hand, higher-order
states featuring two or more nodes are never fully stable, spontaneously
transforming into modes of lower orders. The instability-driven evolution
leads to considerable power losses, caused by the use of the edge absorbers
in the direct simulations, at higher values of LI, and to negligible losses
at lower values. The numerical findings presented in this study suggest new
possibilities for experimental realization of the predicted phenomenology.

A natural direction for the extension of the analysis is to develop it in
the two-dimensional geometry with fractional diffraction. In that case, new
options will be to construct vortex solitons and study their stability.
\section*{Declaration of competing interest}
The authors declare that they have no known competing financial interests or personal relationships that could have appeared to influence the work reported in this paper.

\section*{Acknowledgments}
M.C.P.S. and W.B.C. thank the financial support of the Brazilian agencies CNPq (\#306105/2022-5) and CAPES. This work was also performed as part of the Brazilian National Institute of Science and Technology (INCT) for Quantum Information (\#465469/2014-0). The work of B.A.M. was supported, in part, by grant No. 1695/22 from the Israel Science Foundation.
 \bibliographystyle{elsarticle-num} 
 \bibliography{Refs}





% \end{thebibliography}
\end{document}